# FORMALIZATION OF THE DATA FLOW DIAGRAM RULES FOR CONSISTENCY CHECK

Rosziati Ibrahim and Siow Yen Yen

Department of Software Engineering, Faculty of Computer Science and Information Technology, Universiti Tun Hussein Onn Malaysia (UTHM),
Parit Raja, 86400, Batu Pahat, Johor Malaysia
rosziati@uthm.edu.my
yenyen0831@hotmail.com

## ABSTRACT

*In system development life cycle (SDLC), a system model can be developed using Data Flow Diagram (DFD). DFD is graphical diagrams for specifying, constructing and visualizing the model of a system. DFD is used in defining the requirements in a graphical view. In this paper, we focus on DFD and its rules for drawing and defining the diagrams. We then formalize these rules and develop the tool based on the formalized rules. The formalized rules for consistency check between the diagrams are used in developing the tool. This is to ensure the syntax for drawing the diagrams is correct and strictly followed. The tool automates the process of manual consistency check between data flow diagrams.*

## 1. INTRODUCTION

System development life cycle (SDLC) is an essential process uses during the development of any system. SDLC consists of four main phases. They are planning, analysis, design and implementation. During analysis phase, context diagram and data flow diagrams are used to produce the process model of a system. A consistency of the context diagram to lower-level data flow diagrams is very important in smoothing up developing the process model of a system. However, manual consistency check from context diagram to lower-level data flow diagrams using a checklist is time-consuming process [1]. At the same time, the limitation of human ability to validate the errors is one of the factors that influence the correctness and balancing of the diagrams [2]. This paper presents a technique for modeling data flow diagram rules and proposes a formalization of its rules. The tool is then developed based on the formalized rules. The purpose for the development of a tool is to automate the manual consistency check between data flow diagrams (DFDs) based on the rules of DFDs. The tool serves two purposes: as an editor to draw the diagrams and as a checker to check the correctness of the diagrams drawn as well as consistency between diagrams. The consistency check from context diagram to lower-level data flow diagrams is automated to overcome the manual checking problem.

The motivation of formalizing the rules of data flow diagrams is because DFD has been used in a widely basis for modeling any system but still lacking a precise understanding. Therefore, by formalizing the DFD rules, we can get a formal model of DFD rules. This formal model can be used to ensure that the diagrams drawn are correct and they are consistent with each other.

The rest of this paper is organized as follows. The review of DFD is in Section 2 and the discussion on the related works is in Section 3. Section 4 discusses the syntax and semantics rules of DFD and Section 5 formalizes the DFD rules. The development of the tool is discussed in Section 6 and finally, Section 7 concludes the paper.

## 2. OVERVIEW OF DFD

SDLC is a process uses during the development of software system starting from planning until the implementation phase. Data flow diagramming, on the other hand, is used to produce the process model during the analysis phase [3]. Process model is very important in defining the requirements in a graphical view. Therefore, the reliability of the process model is the key element to improve the performance of the following phases in SDLC.

SDLC is also used to understand on how an information system can support business needs, designing the system, building the system and delivering the system to users [3]. SDLC consists of four fundamental phases, which are analysis, design, implement and testing phases. In the analysis phase, requirements of a system are identified and refined into a process model. Process model can be used to represent the processes or activities that are performed in a system and show the way of data moves among the processes. In order to diagram a process model, data flow diagramming is needed. Dixit *et al.* [4] define data flow diagram as a graphical tool that allows system analysts and users to depict the flow of data in an information system.

Normally, the system can be physical or logical, manual or computer based. Data flow diagram symbols consist of four symbols which are processes, data flows, data stores and external entities. The standard set of symbols that will be used in this paper is devised by Gane and Sarson symbols in [3]. Table 1 shows these symbols.

Table 1 : Symbols for DFD elements in [3]

| Symbol | Element Name |
|---|---|
| 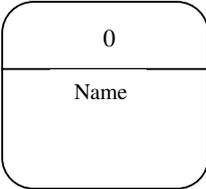 | Process |
| 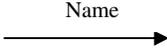 | Data Flow |
| 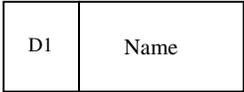 | Data Store |
| 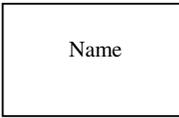 | External Entity |

In data flow diagram, the highest-level view of the system is known as context diagram. The next level of data flow diagram is called the level 0 data flow diagram which represents a system's major processes, data flows and data stores at a high level of detail. Every process in the level n-1 data flow diagram would be decomposed into its lower-level data flow diagram which is level n data flow diagram. The key principle in data flow diagram is to ensure balancing which means that the data flow diagram at one level is accurately represented in the next level data flow diagram when developing a project. The ideal level of decomposition is to decompose the system until system analysts and users can provide a detailed description of the process whereby the process descriptions is not more than one page. The final set of data flow diagrams is validated for ensuring quality. In general, there are two types of problems that can occur in data flow diagrams which are syntax errors and semantics errors. Semantics errors are more complicated than syntax errors due to a set of rules that need to be followed in order to identify them. For example, every process has at least one input data flow and every process has at least one output data flow. Therefore, understanding the set of rules for data flow diagrams is important. Once the rules are understand, a tool can be developed based on the rules so that the tool can perform consistency check between context diagram to level 0 data flow diagram. The tool is also able to perform grammatical errors checking within or across data flow diagrams in order to achieve consistency. By using this tool, the correctness and reliability of data flow diagrams can be increased.

## 3. RELATED WORK

According to Lucas *et al.* [2], consistency problems have existed in Information System development since its beginning and are usually linked to the existence of multiple models or views which participate in the development process. Tao and Kong [5] state that a data flow diagram is visual and informal, hence, it is easy to learn and use. However, its informality makes it difficult to conduct formal verification of the consistency and completeness of a data flow diagram specification.

Dixit *et al.* [4], on the other hand, defined data flow diagram consistency is the extent to which information contained on one level of a set of nested data flow diagram is also included on other levels. According to Tao and Kong [5], the child data flow diagram that results from decomposition is consistent with the precedence relation for the parent process and does not introduce additional precedence relationships between the input and output flows of the parent process. Recently, many systems have been developed to provide automatic support for data flow diagrams specifications. However, all of these systems are lack the ability to manipulate the semantics of data flow diagram specification [6]. Research done by Lee and Tan [7] cover the modelling of DFD using Petri Net model. In their research, they check consistency of the DFDs by enforcing constraints on their Petri Net model. Tong and Tang [6], on the other hand, model the DFD using temporal logic language. A method for checking consistency for Unified Modelling Language (UML) specification, on the other hand, has been done for example in [2], [8] and [9].

Various researches also stated that no formal language has been currently used for semantic specification of data flow diagram ([1], [6], and [10]). However, Tao and Kong [6] point out that there are few development environments or CASE tools provide automated verification facilities that can detect inconsistency and incompleteness in a data flow diagram specification. Dixit *et al.* [4] therefore describe that the concept of data flow diagram consistency is refers to whether or not the depiction of the system shown at one level of a nested set of data flow diagram is compatible with the depictions of the system shown at other levels. They also state that a consistency check facility with a CASE tool will be helpful for the practitioners. Consistency in process decomposition, on the other hand, means that the precedence relation is faithfully inherited by the child data flow diagram [5]. Ahmed Jilani *et al.* [1], on the other hand, state that notations used in the data flow diagram are usually graphical and different tools and practitioners interpret their notations differently. Therefore, a well-defined semantics or data flow diagram formalism could help to reduce inconsistencies and confusion.

This paper formalizes important DFD rules to address the consistency issues in DFD. Our research focuses on consistency check between data flow diagrams and develops a tool to automate a consistency check between data flow diagrams based on the formal notations used for the DFD rules.

## 4. SYNTAX AND SEMANTIC RULES OF DFD

Data flow diagrams are illustrated movement of data between external entities and the processes and data stores within a system [11]. According to Donald and Le Vie [12], data flow diagrams are a tool that can reveal relationships among and between the various components in a program or system. Tao and Kong [5], on the other hand, stated that data flow diagram technique is effective for expressing functional requirements for large complex systems. Definition 1 gives the definition of data flow diagram where there are four symbols in the data flow diagram which are processes, data flows, data stores and external entities (source/sink). In general, there are two commonly used styles of symbols in data flow diagram as described in [3] and [4]. For our research, we will use Gane and Sarson symbols as described in [3] which appear in Table 1.

*Definition 1*: A Data Flow Diagram consists of:
- Processes
- Data Flows
- Data Stores
- External Entites

where
 - A process is an activity or a function that is performed for some specific business reason;
 - A data flow is a single piece of data or a logical collection of several pieces of information;
 - A data store is a collection of data that is stored in some way;
 - An external entity is a person, organization, or system that is external to the system but interact with it.

The highest-level of data flow diagram is known as the context diagram. According to Jeffrey *et al.* [8], a context diagram is a data flow diagram of the 10 scope of an organizational system that shows the system boundaries, external entities that interact with the system and the major information flows between the entities and the system. Dennis *et al.* [3] state that the context diagram shows the overall business process as just one process and shows the data flows to and from external entities. Data stores are not usually included on the context diagram. The context diagram therefore is decomposed into the lower-level diagram which is level 0 data flow diagram. In fact, each process on the level 0 data flow diagram can be decomposed into more explicit data flow diagram, called level 1 diagram and can be further decomposed into next lower-level diagram when it is needed. In general, there are two fundamentally different types of problems that can occur in data flow diagrams which are syntax errors and semantics errors. Tao and Kong [5] defined the syntax of the data flow diagram is how components are interconnected through data flows and what components constitute the subsystem being modeled. The semantics of the data flow diagram, on the other hand, is how data flows are interrelated in terms of data transformations. Dennis *et al.* [3] claimed that syntax errors are easier to find and fix than are semantics errors because there are clear rules that can be used to identify them. There is a set of rules that must be followed by analysts when drawing data flow diagrams in order to evaluate data flow diagrams for correctness [12]. Definition 2 until Definition 8 stated these rules.

*Definition 2*: Rules of data flow diagrams:
- At least one input or output data flow for external entity
- At least one input data flow and/or at least one output data flow for a process
- Output data flows usually have different names than input data flows for a process
- Data flows only in one direction
- Every data flow connects to at least one process

*Definition 3*: Unique name in data flow diagrams:
- A unique name (verb phase), a number and a description for a process
- A unique name that is a noun and a description for a data flow
- A unique name that is a noun and a description for data store
- A unique name that is a noun and a description for external entity

*Definition 4*: Consistency:
- Every set of data flow diagrams must have one context diagram.

*Definition 5*: Consistency Viewpoint:
- There is a consistency viewpoint for the entire set of DFDs.

*Definition 6*: Decomposition:
- Every process is wholly and completely described by the processes on its children DFDs.

*Definition 7*: Balancing:
- Every data flow, data store and external entity on a higher level DFD is shown on the lower-level DFD that decomposes it.

*Definition 8*: Data Store:
- For every data store, data cannot move directly from one data store to another data store.
- Data must be moved by a process.

Definitions 2 until 8 explain the fundamental rules of data flow diagrams. The consistency between context diagram and data flow diagram is very important and the rules for these consistency is captured in Definitions 4 and 5. Following on the consistency issue, Definition 6 addresses aspect on decomposition of the processes to its lower level of DFD and Definition 7 addresses aspect of balancing of DFD elements to its lower level of DFD. Syntax rules are used to verify syntax errors within the DFD. The syntax rules are defined in Definition 9.

*Definition 9*: Syntax rules of data flow diagram:
- At least one input data flow for a process
- At least one output data flow for a process
- Process from external entity cannot move directly to another external entity
- At least one input data flow for a data store
- At least one output data flow for a data store
- Data from one data store cannot move directly to another data store

Based on Definition 9, six syntax rules are used in order to verify the correctness of the context diagram and level 0 data flow diagram. However, the syntax rules of data store only applied in level 0 data flow diagram. Semantics rules are used to verify semantics errors from context diagram to level 0 data flow diagram. The semantics rules are defined in Definition 10.

*Definition 10*: Semantic rules of data flow diagram:
- The total number and name of external entities in context diagram are the same as in level 0 DFD
- The total number and name of data flows between process and external entity in context diagram are same as level 0 DFD
- The total number and name of external entities in level 0 DFD are same as context diagram
- The total number and name of data flows between process and external entity in level 0 DFD are the same as in context diagram

The semantics rules defined in Definition 10 are used to perform consistency check from context diagram to level 0 data flow diagram. We then formalize the DFD rules and represented them using mathematical notations in order to better understand the rules. Similar approach for formalization of DFDs is in [6] and [7], where Tong and Tang [6] use temporal logic language and Lee and Tan [7] use Petri Net model. Gao and Huaikou [13], on the other hand, integrate structured approach with object-oriented approach and suggest a formal language using Z notations for predicate data flow diagram (PDFD). The formalization of DFD rules is discussed in next section.

## 5. FORMALIZATION OF DFD RULES

This section formalizes the important DFD rules based on the syntax and semantics rules of DFD given in Section 4. The mathematical notations are used, for better understanding of the DFD rules.

*Rule 1*. Let *D* be a data flow diagram, then
$$D = \{P, F, S, E\} \tag{1}$$
where
$P = \{p_1, p_2, p_3..., p_m\}$ is a finite set of processes;
$F = \{f_1, f_2, f_3..., f_m\}$ is a finite set of data flows;
$S = \{s_1, s_2, s_3..., s_m\}$ is a finite set of data stores;
$E = \{e_1, e_2, e_3..., e_m\}$ is a finite set of external entities;

Rule 1 defines the data flow diagram. Data flow diagram consists of a set of processes, data flows, data stores and external entities.

*Rule 2*. Let *C* be a context diagram then
$$C = \{<e_i, f_j, p_1>, <p_1, f_k, e_i>\}, j \neq k, 1 \leq i, j, k \leq m \tag{2}$$

Rule 2 defines the context diagram. Context diagram consists of one process only and a set of external entities and data flows. Data flow can be connected from external entity to a process and vice versa but the data flow must be a different data flow. Note that, data store can only exist in data flow diagram but not context diagram.

*Rule 3.*

$$\forall p_i, p_j \in P, p_i \neq p_j, 1 \leq i, j \leq m \qquad (3)$$

From Rule 3, the name of the process is unique. For any process, no duplication is allowed. The same rules apply for data flows (Rule 4), data stores (Rule 5) and external entities (Rule 6). The name is unique and duplication of name is not allowed.

*Rule 4.*
$$\forall f_i, f_j \in F, f_i \neq f_j, 1 \leq i, j \leq m \qquad (4)$$

*Rule 5.*
$$\forall s_i, s_j \in S, s_i \neq s_j, 1 \leq i, j \leq m \qquad (5)$$

*Rule 6.*
$$\forall e_i, e_j \in E, e_i \neq e_j, 1 \leq i, j \leq m \qquad (6)$$

*Rule 7.*
$$\forall f_i \in C, \exists f_j \in D, f_i = f_j, 1 \leq i, j \leq m \qquad (7)$$

Rule 7 indicates that for any data flow that belongs to context diagram, that data flow must exist in data flow diagram. The same rule applies to external entity. That is, for any external entity that belongs to context diagram, that external entity must exist in data flow diagram (Rule 8).

*Rule 8.*
$$\forall e_i \in C, \exists e_j \in D, e_i = e_j, 1 \leq i, j \leq m \qquad (8)$$

*Rule 9.*
$$\forall e_i, e_j, s_i, s_j, f_k \in D \text{ , then} \qquad (9)$$
$$D \neq \{<e_i, f_k, e_j>\} \text{ and } D \neq \{<s_i, f_k, s_j>\}, 1 \leq i, j, k \leq m$$

Rule 9 indicates that for any data flow diagram, a data flow cannot connect from one external entity to another external entity and a data flow cannot also connect from one data store to another data store.

*Rule 10.*
$$\forall e_i, s_j, f_k \in D \text{ , then} \qquad (10)$$
$$D \neq \{<e_i, f_k, s_j>\} \text{ and } D \neq \{<s_j, f_k, e_i>\}, 1 \leq i, j, k \leq m$$

Rule 10 indicates that for any data flow diagram, a data flow cannot connect from one external entity to data store and a data flow cannot also connect from one data store to external entity.

*Rule 11.*
$$\forall p_i, s_i, f_j, f_k \in D \text{ , then} \qquad (11)$$
$$D = \{<p_i, f_j, s_i>\} \text{ and } D = \{<s_i, f_k, p_i>\},, f_j \neq f_k, 1 \leq i, j, k \leq m$$

Rule 11 indicates that for any data flow that connect from process to data store, another data flow can connect from data store to process but must be different from the previous used data flow.

*Rule 12.*
$$\forall e_i, p_i, f_j, f_k \in D, \text{ then} \tag{12}$$
$$D = \{<e_i, f_j, p_i>\} \text{ and } D = \{<p_i, f_k, e_i>\},, f_j \neq f_k, 1 \leq i, j, k \leq m$$

Rule 12 indicates that for any data flow that connect from external entity to process, another data flow can connect from process to external entity but must be different from the previous used data flow.

*Rule 13.*
$$\forall p_i, p_j, f_j, f_k \in D, \text{ then} \tag{13}$$
$$D = \{<p_i, f_j, p_j>\} \text{ and } D = \{<p_j, f_k, p_i>\},, f_j \neq f_k, 1 \leq i, j, k \leq m$$

Rule 13 indicates that for any data flow that connect from one process to another process, another data flow can connect from another process to previous process but must be different from the previous used data flow.

## 6. THE TOOL

The tool is developed based on the set of rules imposed by data flow diagrams as described in Section 5. A graphical layout is used in order to use the tool as an editor for drawing the diagrams and as a checker as well to check the correctness of the diagrams. Figure 1 shows the main interface of the tool.

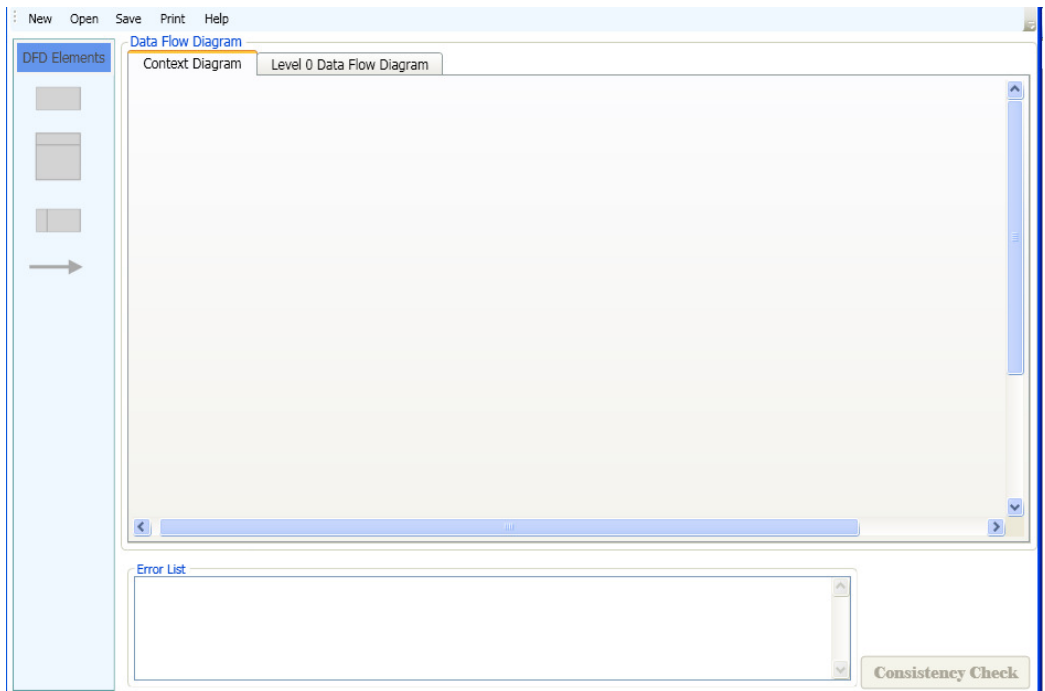

Figure 1. Interface of the Tool

From Figure 1, the main interface provides a platform that allows the user to input both diagrams by using the data flow diagram elements provided. The main interface includes four main parts where the top of the interface is the menu bar consists of five menu functions, the toolbar of the data flow diagram elements is in the left-side of the interface, the bottom-right is an error list text box and a "Consistency Check" button. The rest of interface is the drawing panel for user to draw the particular diagram. The five functions in menu bar are open a new file, open a saved file, save the data flow diagrams, print the data flow diagrams and open a help menu. In the toolbar, there are four data flow diagram elements which are process, external entity, data flow and data store. User is allowed to drag and drop the data flow diagram elements on the drawing panel. "Consistency Check" button, on the other hand, is used to perform the consistency check after both diagrams are created. Therefore, the tool serves two purposes. The first purpose is as an editor for drawing the context diagram and level 0 data flow diagram and the second purpose is as a checker for checking the consistency between context diagram and level 0 data flow diagram.

In this paper, we give one simple example of an academic information system and use the tool to represent the context diagram and its level 0 of data flow diagram. Figure 2 shows the example of the context diagram for a lecturer who is going to use the Academic Information System (AIS). A lecturer can send his or her academic information to the system and can get a list of academicians from the system.

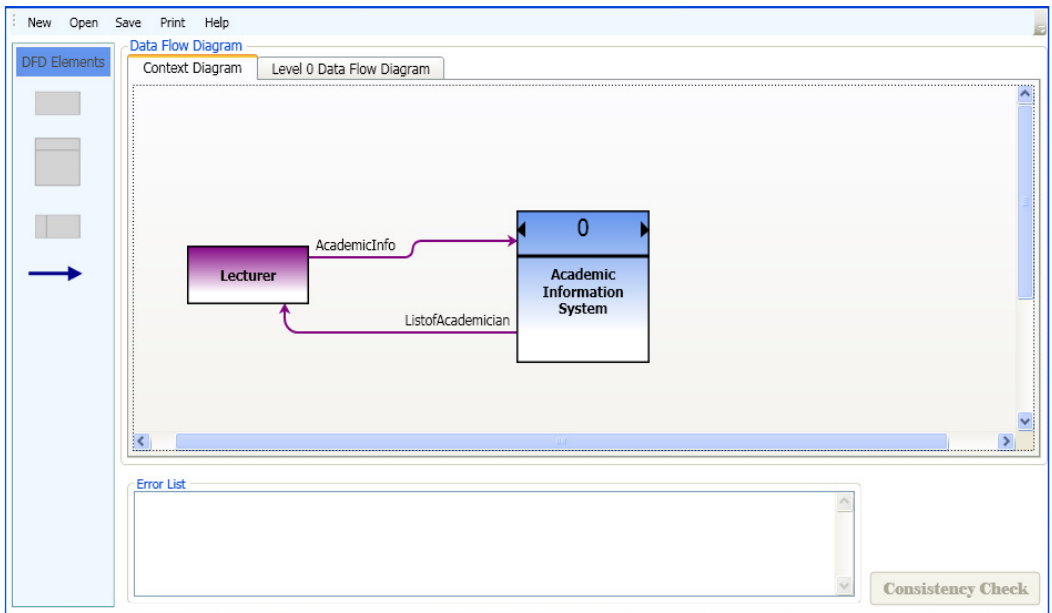

Figure 2. Example of Context Diagram

In reference to Figure 2, using Rule 2, the AIS consists of 1 entity, 1 process and 2 data flows. That is, $C = \{<e_1, f_1, p_1>, <p_1, f_2, e_1>\}$. The context diagram also follows Rule 4 for the uniqueness of data flow name. From context diagram, a level 0 data flow diagram can be drawn as shown in Figure 3.

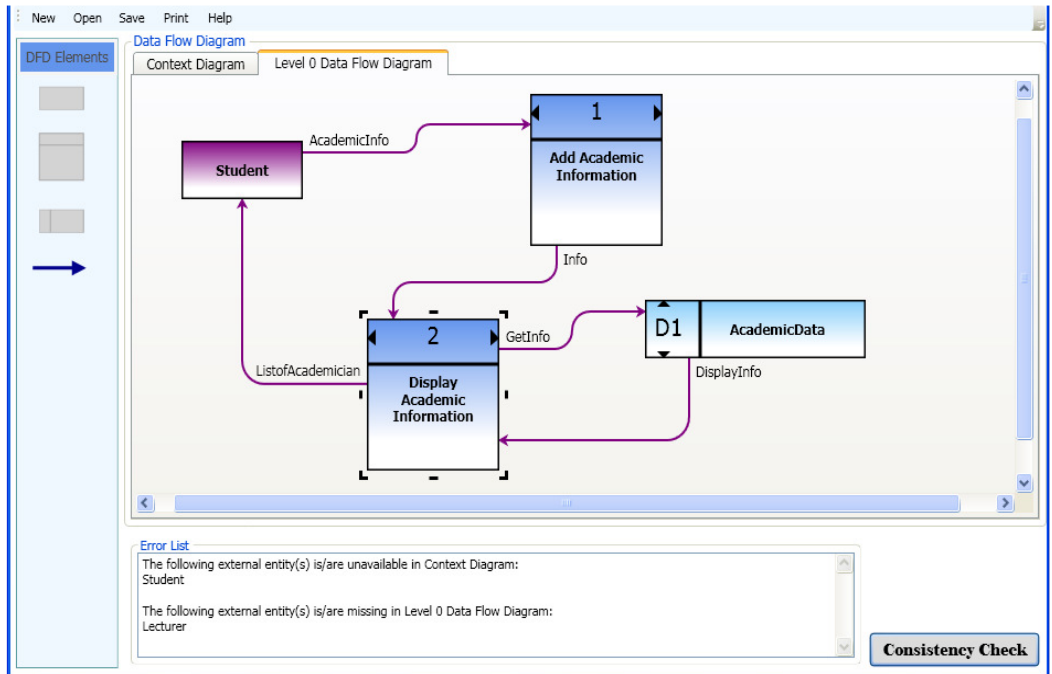

Figure 3. Example of Level 0 Data Flow Diagram

Based on Figure 3, level 0 data flow diagram for AIS consists of 2 processes, 1 external entity, 1 data store and 5 data flows. That is,

$$D = \{<e_2, f_1, p_2>, <p_3, f_2, e_2>, <p_2, f_3, p_3>, <p_3, f_4, s_1>, <s_1, f_5, p_3>\}$$

Data flow diagram follows Rules 3 until 6 for uniqueness name used. For data flow, Rule 7 is also followed. However, for external entity, Rule 8 is not followed. That is, if there exist external entity in context diagram, that external entity must exist in data flow diagram. Therefore, the data flow diagram consists of syntax errors.

The tool verifies the syntax errors for all the data flow diagram elements used. When there is any syntax errors exist in the data flow diagram elements, the tool displays an error message to user. Figure 3 shows an existence of syntax error from the data flow diagram. Since the entity from context diagram is Lecturer, the tool informs the inconsistency between context diagram and data flow diagram. The user can then use the editor of the tool to correct the syntax error. If the entity is correct, the tool validates the consistency check as shown in Figure 4.

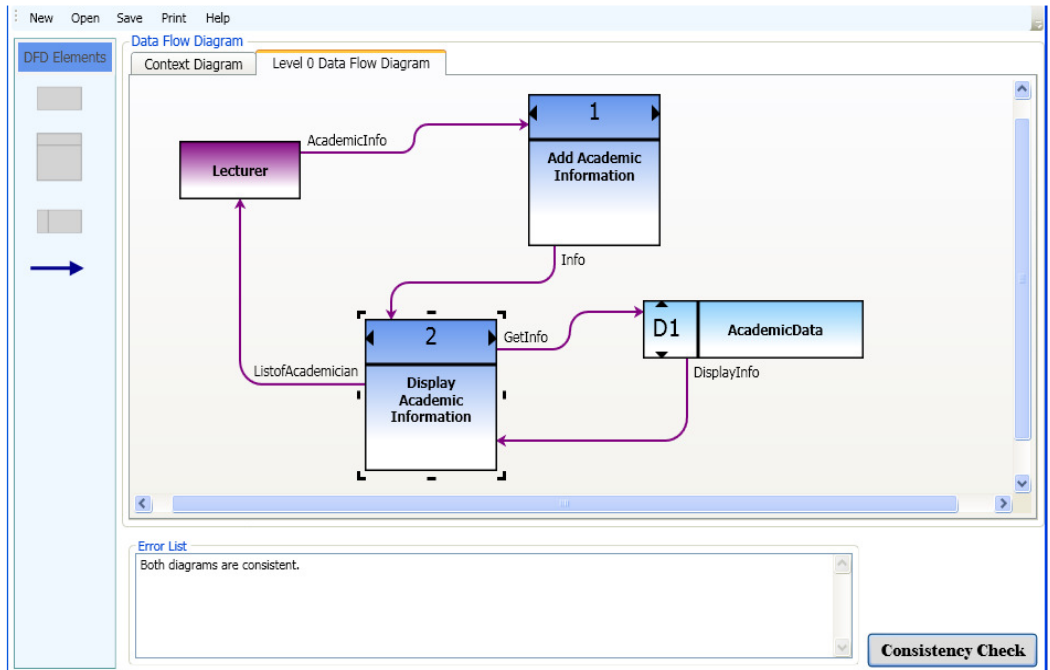

Figure 4. Consistency Check

Based on Figure 4, level 0 data flow diagram consists of the correct external entity. That is,

$$D = \{<e_1, f_1, p_2>, <p_3, f_2, e_1>, <p_2, f_3, p_3>, <p_3, f_4, s_1>, <s_1, f_5, p_3>\}$$

Data flow diagram described in Figure 4 follows Rule 8. That is the external entity exists in context diagram as well as in data flow diagram. The tool verifies that both diagrams are consistent.

Once the diagrams are drawn, they can be saved to a new or existing folder. The tool allows the user to save and print the diagrams. The user can open the folder again for viewing or editing of the diagrams. The user can also print the diagrams. The Help menu can be used for getting more information regarding the tool.

The development of the tool also incorporated the rules for syntax errors. Rules 9 until 12 are also implemented inside the tool. For example, if a user wants to draw a data flow from one external entity to another external entity, the tool prompts a syntax error indicating that such connection cannot be done. Figure 5 shows the example of it.

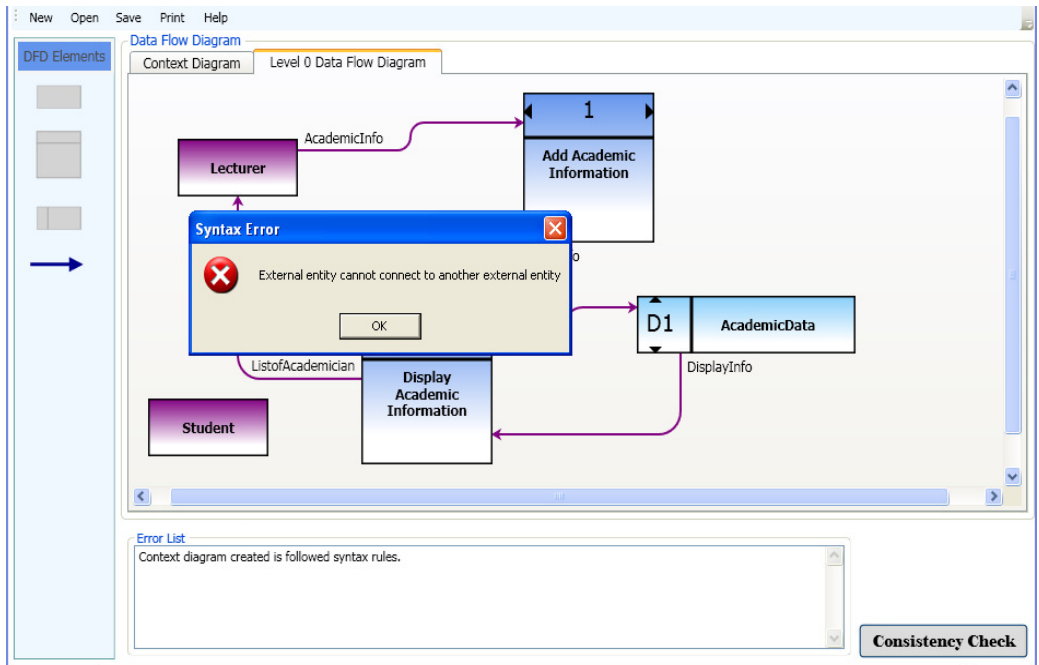

Figure 5. Example of Syntax Error for Rule 9.

Figure 5 shows that if a user wants to connect a data flow from one external entity to another external entity, the tool prompts a syntax error indicating that the drawing cannot be done. The same goes to other rules such as Rule 10. The tool prompts a syntax error if a user tries to connect a data flow from external entity to data store.

We demonstrate another example of using our tool to represent the context diagram and its level 0 of data flow diagram. Figure 6 shows the example of the context diagram for a student who wants to borrow book using the Library System (LS). A Student can send his or her list of book to borrow to the system and can get a list of book that is available for him or her to borrow from the system.

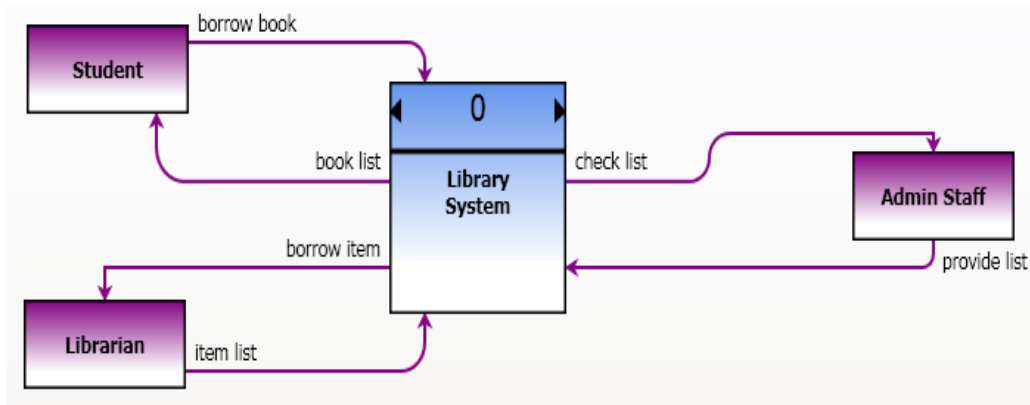

Figure 6. Context Diagram for Library System (LS)

International Journal of Software Engineering & Applications (IJSEA), Vol.1, No.4, October 2010

In reference to Figure 6, using Rule 2, the LS consists of 1 process, 3 entities and 6 data flows. The context diagram also follows Rule 4 for the uniqueness of data flow name. From context diagram, a level 0 data flow diagram can be drawn as shown in Figure 7.

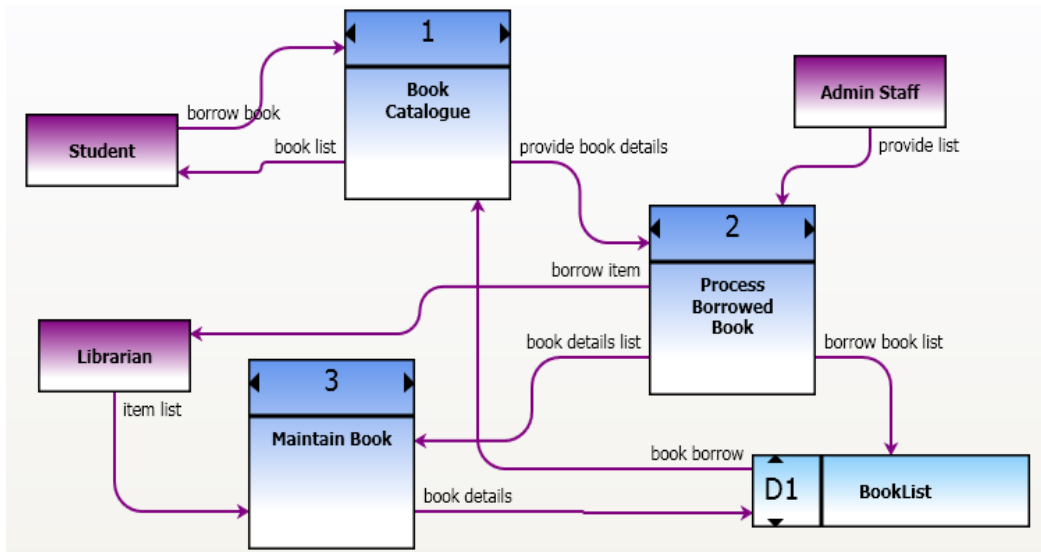

Figure 7. Example of Level 0 Data Flow Diagram

Based on Figure 7, level 0 data flow diagram for LS consists of 3 processes, 3 entities, 1 data store and 10 data flows. Data flow diagram follows Rules 3 until 6 for uniqueness name used. For external entity, Rule 8 is followed. However, for data flow, Rule 7 is not followed. That is, if there exist data flow in context diagram, that data flow must exist in data flow diagram. Therefore, the data flow diagram consists of syntax errors.

The tool validates the syntax errors for all the data flow diagram elements used. When there is any syntax errors exist in the data flow diagram elements, the tool displays an error message to user. Figure 7 shows an existence of syntax error from the data flow diagram. Since the data flow from context diagram is <check list>, the tool informs the inconsistency between context diagram and data flow diagram. The user can then use the editor of the tool to correct the syntax error. If the data flow is correct, the tool can also be used for the consistency check. Figure 8 shows that the diagrams (context diagram and level 0 data flow diagrams) are consistent.

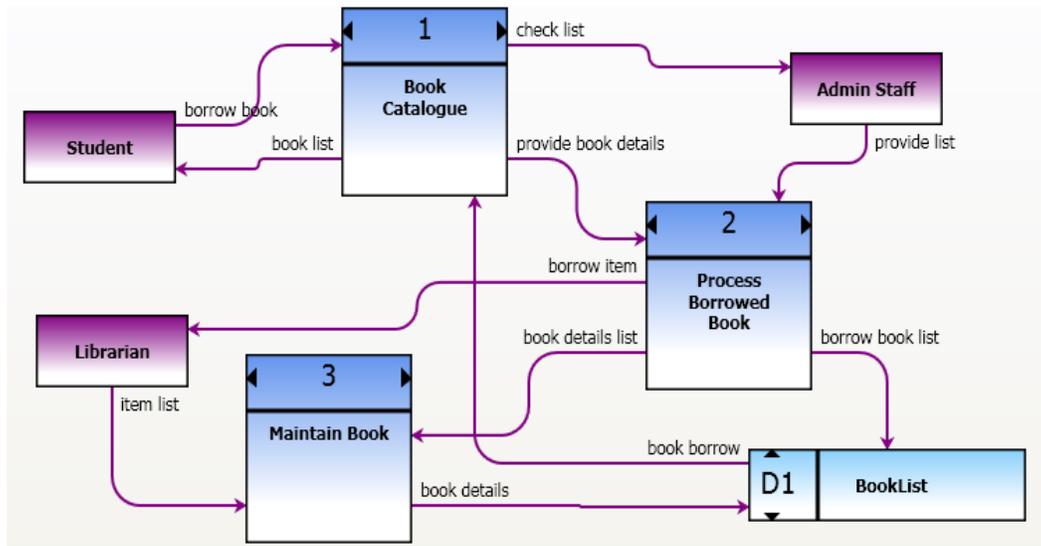

Figure 8. Consistency Check

Based on Figure 8, level 0 data flow diagram consists of the correct data flow. Data flow diagram follows Rule 7. That is the data flow exists in context diagram as well as in data flow diagram. The tool will verify that both diagrams are consistent.

Once the diagrams are drawn, they can be saved to a new or existing folder. The tool allows the user to save and print the diagrams. The user can open the folder again for viewing or editing of the diagrams. The user can also print the diagrams. The Help menu can be used for getting more information regarding the tool.

The development of the tool also incorporated the rules for syntax errors. Rules 9 until 12 are also implemented inside the tool. For example, if a user wants to draw a data flow from one external entity to data store, the tool prompts a syntax error indicating that such connection cannot be done. Figure 9 shows the example of it.

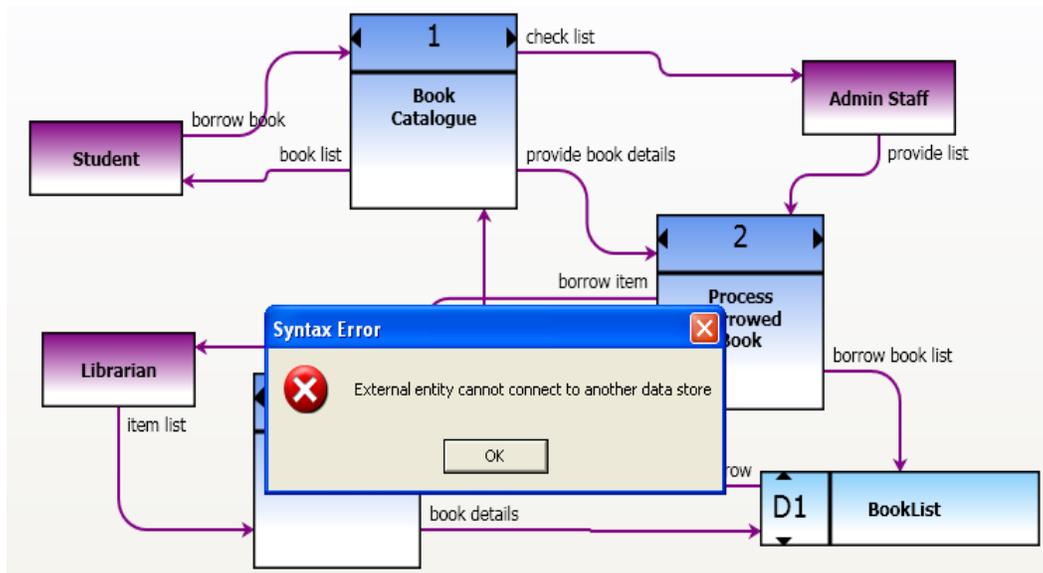

Figure 9. Example of Syntax Error for Rule 10.

Figure 9 shows that if a user wants to connect a data flow from one external entity to data store, the tool prompts a syntax error indicating that the drawing cannot be done. The same goes to other rules such as Definition 9. The tool prompts a syntax error if a user tries to connect a data flow from external entity to data store. The tool ensures the correctness of the diagrams drawn and the balancing of every data flow and external entity between context diagram with level 0 data flow diagram.

## 7. CONCLUSIONS

This paper has discussed how to model a business process flow using data flow diagrams and presented a set of syntax and semantics rules of data flow diagrams. The rules are then being formalized and used to automate the process of checking the consistency between the context diagram and level 0 data flow diagrams. The automatic checking of consistency overcomes the time-consuming process of manually checking the correctness of the diagrams. The developers can use the tool for drawing and designing their process model of the system that they want to develop.

The tool serves two purposes. The first purpose is as an editor to draw the diagrams and the second purpose is as a checker to check the correctness of the diagrams drawn as well as consistency between the diagrams. Our tool has several advantages. First, we can minimize the syntax errors when drawing the diagrams since the tool prevents the user from making such errors. Second, the correctness of diagrams is guaranteed since the consistency check between diagrams are also done via the tool.

## ACKNOWLEDGEMENTS

This research is supported by the Science Fund under Ministry of Science, Technology and Innovation (MOSTI), Malaysia.

**Authors**

Rosziati Ibrahim is with the Software Engineering Department, Faculty of Computer Science and Information Technology, Universiti Tun Hussein Onn Malaysia (UTHM). She obtained her PhD in Software Specification from the Queensland University of Technology (QUT), Brisbane and her MSc and BSc (Hons) in Computer Science and Mathematics from the University of Adelaide, Australia. Her research area is in Software Engineering that covers Software Specification, Software Testing, Operational Semantics, Formal Methods, Data Mining and Object-Oriented Technology.

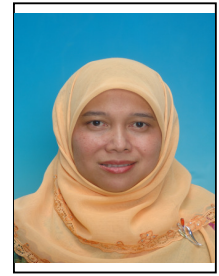

Siow Yen Yen is a student at the Department of Software Engineering, Faculty of Computer Science and Information Technology, Universiti Tun Hussein Onn Malaysia (UTHM), Batu Pahat, Johor, Malaysia.

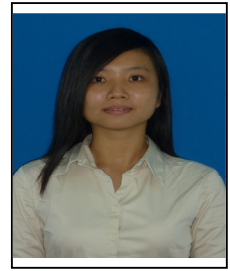